\documentclass[singlecolumn,11pt]{article}
\usepackage{spconf}
\usepackage{amsmath,graphicx}

\usepackage{amsmath,amssymb}
\usepackage[colorlinks=true,bookmarks=false,citecolor=blue,urlcolor=blue]{hyperref} 
\usepackage{graphicx}
\usepackage[english]{babel}
\usepackage[T1]{fontenc}  
\usepackage[latin1]{inputenc} 
\usepackage{url}
\usepackage{graphicx}
\usepackage{subfigure}
\usepackage{xcolor}
\usepackage{amsmath,amsfonts,amssymb}
\usepackage{booktabs}
\usepackage{epsfig}
\usepackage{pstricks}
\usepackage{pst-node}
\usepackage{pst-grad}
\usepackage{ifthen}
\usepackage{algorithm}
\usepackage{algorithmic}	
\usepackage{setspace}           
\usepackage{epstopdf}	
\usepackage{enumerate} 
\usepackage{hyperref}
\usepackage{multicol} 
\usepackage{enumitem} 
\usepackage{cite} 
\usepackage{verbatim} 
\usepackage{siunitx} 
\usepackage{fancyhdr}

\usepackage{tcolorbox}
\usepackage{framed}
\usepackage{caption}
\allowdisplaybreaks

\newcommand{\mypar}[1]{\bigskip\noindent {\bf #1}}

\definecolor{red}{RGB}{153,0,0}
\newcommand{\bx}{\mathbf{x}}

\newcommand{\bz}{\mathbf{z}}
\newcommand{\bX}{\mathbf{X}}

\newcommand{\bp}{\mathbf{p}}

\newcommand{\bd}{\mathbf{d}}

\newcommand*{\dif}{\mathop{}\!\mathrm{d}}

\title{Light-Field Microscopy for optical imaging of neuronal activity: when model-based methods meet data-driven approaches}

\name{
	Pingfan~Song, \;
	Herman Verinaz Jadan, \;
	Carmel L. Howe, \;
	Amanda J. Foust, \;
	Pier Luigi Dragotti%
	\thanks{This work was supported by the Biotechnology and Biological Sciences Research Council (BBSRC grant: BB/R009007/1), Wellcome Trust Seed Award (201964/Z/16/Z), Royal Academy of Engineering Research Fellowship (RF1415/14/26), Engineering and Physical Sciences Research Council (EPSRC grant:EP/L016737/1).}
	\thanks{Pingfan Song is with the Department of Engineering, University of Cambridge, Cambridge, CB2 1PZ, UK. (e-mail: ps898@cam.ac.uk)}
	\thanks{Herman Verinaz Jadan and Pier Luigi Dragotti are with the Department of Electronic and Electrical Engineering, Imperial College London, London, SW7 2AZ, UK. (e-mail: herman.verinazjadan17@imperial.ac.uk, p.dragotti@imperial.ac.uk)}
	\thanks{Carmel L. Howe and Amanda J. Foust are with the Department of Bioengineering, and Center for Neurotechnology, Imperial College London, London, SW7 2AZ, UK. (e-mail: carmel.howe@imperial.ac.uk, a.foust@imperial.ac.uk)}
}

\begin{document}

\maketitle
\vspace{-2cm}

\begin{abstract}
	Understanding how networks of neurons process information is one of the key challenges in modern neuroscience. A necessary step to achieve this goal is to be able to observe the dynamics of large populations of neurons over a large area of the brain. Light-field microscopy (LFM), a type of scanless microscope, is a particularly attractive candidate for high-speed three-dimensional (3D) imaging. It captures volumetric information in a single snapshot, allowing volumetric imaging at video frame-rates. Specific features of imaging neuronal activity using LFM call for the development of novel machine learning approaches that fully exploit priors embedded in physics and optics models. Signal processing theory and wave-optics theory could play a key role in filling this gap, and contribute to novel computational methods with enhanced interpretability and generalization by integrating model-driven and data-driven approaches. This paper is devoted to a comprehensive survey to state-of-the-art of computational methods for LFM, with a focus on model-based and data-driven approaches.
\end{abstract}


\vspace{-0.2cm}

\section{Introduction}

\vspace{-0.2cm}

One of the key goals of neuroscience is understanding how networks of neurons in the brain process information. Achieving this goal requires the ability to capture the dynamics of large populations of neurons at high speed and resolution over a large area of the brain. Whilst there are many viable techniques for observing brain activity, optical imaging with fluorescent indicators is a popular strategy to record the activity of neurons owing to the high spatial resolution and potential scalability. 

\begin{framed}
	\label{Box:ImagingModalities}
	\singlespacing
	\mypar{Box~1: Sequential versus parallel imaging modalities.}
	
	\begin{center}
		\includegraphics[width = 14cm, 
		]{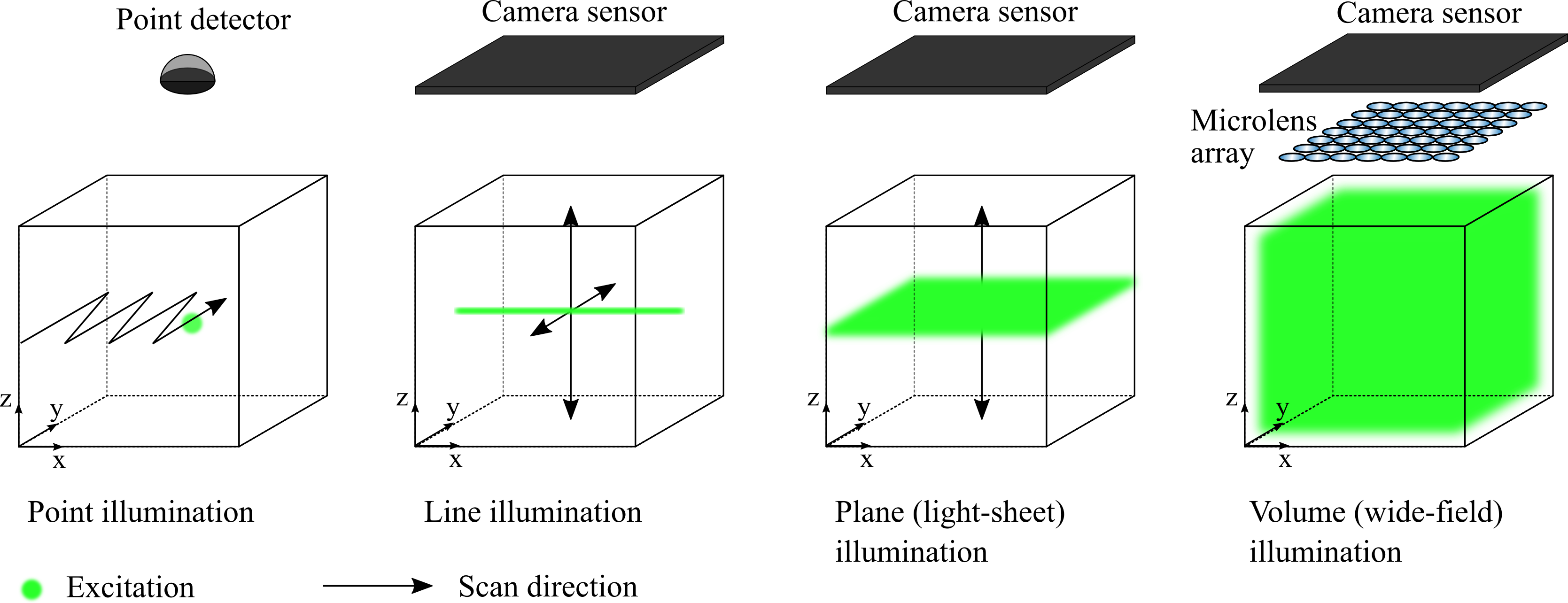}
		\captionof{figure}{Illustration of imaging modalities categorized based on the acquisition mode.
		In a sequential acquisition, as shown in the first sub-figure, a focal spot is scanned across one or more dimensions in space to cover the entire volume. Consequently, the location of the signal is determined by the instantaneous position of the excitation beam, and a point detector can be used to collect emitted fluorescence photons irrespective of the path with which they reach the detector. Such imaging modalities confer robustness to light scattering and are therefore well-suited for deep tissue imaging. This comes at the cost of a reduced temporal resolution. In parallel acquisition modes, as shown in other three sub-figures, some or all voxels are recorded simultaneously on a camera sensor array, which enables higher volume rates. In particular, light-sheet imaging simultaneously scans the light-sheet excitation plane to quickly build up a volume plane by plane. In this way, it enables to image the whole volume plane by plane. In contrast, light-field imaging enables simultaneous recording of all voxels within a volume. Fluorescence generated throughout the volume is captured through a microlens array to encode both position and angular information simultaneously. However, unlike point-scanning modalities, parallel acquisition modes are inherently vulnerable to scatter-induced crosstalk between neighbouring camera pixels. See also~\cite{weisenburger2018guide} for description of different imaging modalities.
		}
		\label{Fig:ImagingModalities}
	\end{center}
\end{framed}

A fluorescent indicator transduces biophysical changes into changes in fluorescence. The most commonly used indicators in functional neuroimaging are those that respond to changes in membrane potential or calcium concentration. Although calcium indicators monitor neuronal membrane potential indirectly, they are most commonly used because of their slow transients and high signal amplitude which make them easier to detect than membrane potential indicators. In particular, due to their slow decay, calcium indicators can be used with diffraction-limited point scanning techniques and single-pixel detectors that collect the emitted fluorescence photons irrespective of the path which they take to reach the detectors, as shown in the leftmost sub-figure in Fig~\ref{Fig:ImagingModalities}. This leads to multi-photon microscopy, an imaging approach robust to scattering and well suited for imaging deep in tissue. However, this approach inherently results in a low temporal resolution due to the use of point scanning techniques, therefore ill-suited to capture fast biological dynamics.

Neurons communicate through electrical impulses lasting around 1 millisecond and repeated at rates up to hundreds of Hertz. Therefore, the study of neural information coding requires high temporal resolution imaging methods. In recent years, we have witnessed substantial progress in fluorescence microscopy imaging in particular due to improved fluorescent indicators of neuronal activity and to the use of alternative scanning strategies for faster acquisition~\cite{schultz2016advances,weisenburger2018guide}. For example, scanning with lines or sheets, instead of points, speeds up acquisition through spatial parallelization (Box~1 and Fig~\ref{Fig:ImagingModalities})~\cite{weisenburger2018guide}. Furthermore, scanless, whole-volume (called "wide-field") is the most efficient illumination scheme, maximizing photon generation rates and imaging speeds. This speed increase, however, comes at the cost of increased background interference since out-of-focus light appears in the in-focus image, reducing contrast. Interestingly, light-field microscopy (LFM) can exploit this out-of-focus fluorescence by reassigning photons back to their correct 3D locations. It becomes a particularly attractive candidate for high-speed three-dimensional (3D) bioimaging.

\begin{figure}[tbh]
	\centering
	\begin{minipage}[b]{1\linewidth}
		\centering
		\includegraphics[width = 16cm, 
		]{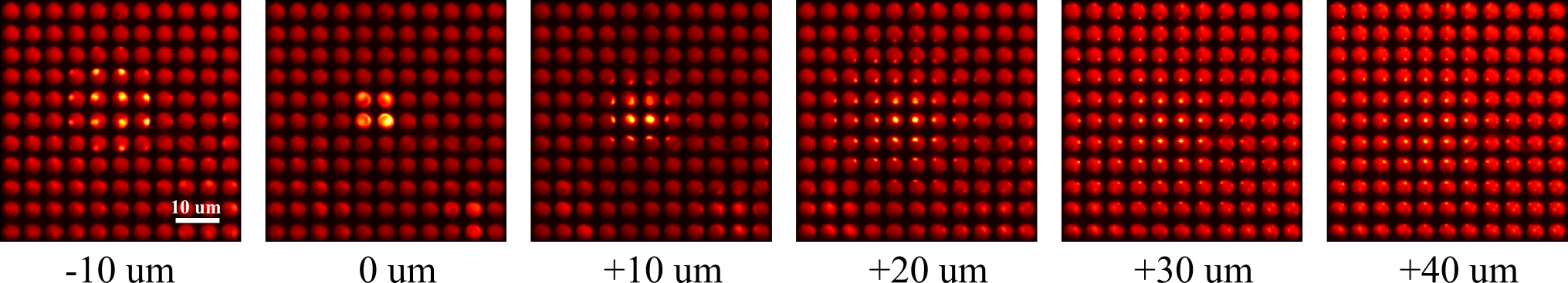}
		\\
		\small (a) Raw LFM data for a neuron at different depths away from the focal plane.
	\end{minipage} 
	\\
	\vspace{+0.2cm}
	\begin{minipage}[b]{0.5\linewidth}
		\centering
		\includegraphics[width = 7.7cm, 
		]{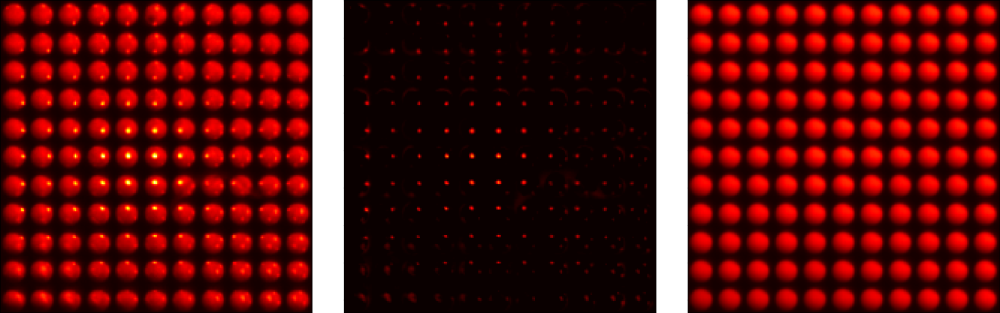}
		\\
		\footnotesize (b)  Light-field for depth 30 \si{\um} (left), foreground (middle) and background (right) separated using matrix factorization.
	\end{minipage} 
	\begin{minipage}[b]{0.48\linewidth}
		\raggedright
		\includegraphics[width = 7.7cm, 
		]{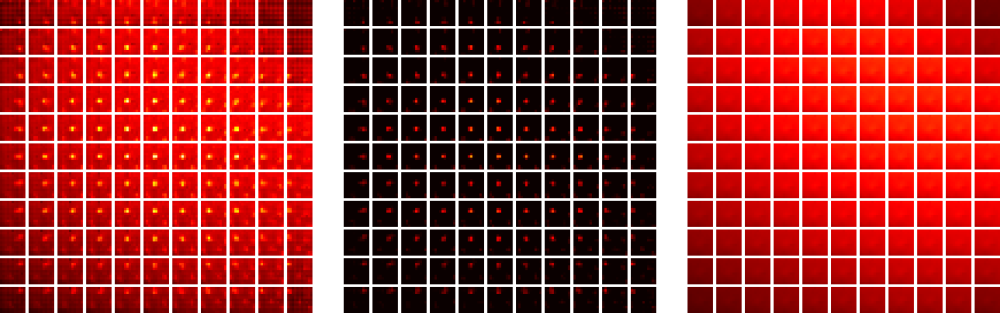}
		\\
		\footnotesize (c)  An array of multi-view images (left) for depth 30 \si{\um} and corresponding foreground (middle) and background (right). 
	\end{minipage} 
	\\
	\vspace{+0.2cm}
	\begin{minipage}[b]{0.99\linewidth}
		\centering
		\includegraphics[width = 14cm, 
		]{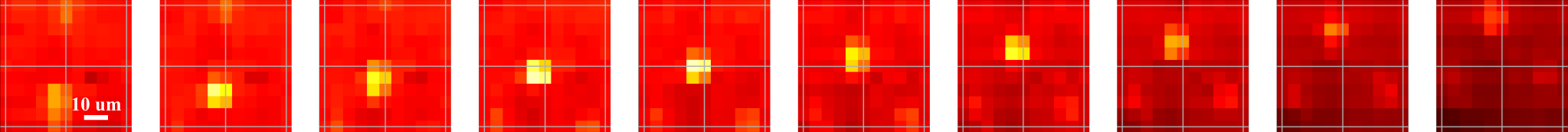}
		\\
		\footnotesize (d) A series of views extracted from the central column of the multi-view array for depth 36 \si{\um}. Note that view changes vertically.
	\end{minipage} 
	\\
	\vspace{+0.2cm}
	\begin{minipage}[b]{0.99\linewidth}
		\centering
		\includegraphics[width = 14cm, 
		]{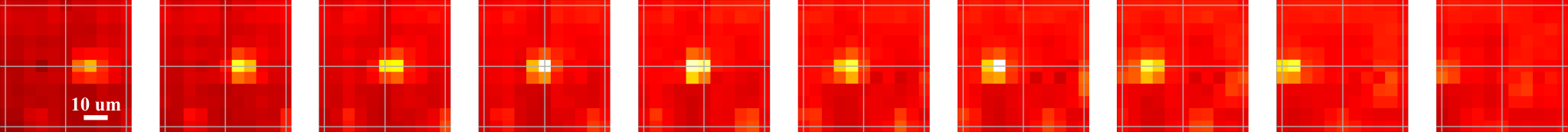}
		\\
		\footnotesize (e) A series of views extracted from the central row of the multi-view array for depth 36 \si{\um}. Note that view changes horizontally.
	\end{minipage} 
	
	\caption{
		(a) Raw LFM images of a mouse neuron (expressing a genetically encoded fluorophore) at different depths away from the focal plane. Due to scattering and blurring, the images have a bright background, and those at deeper depths have a weaker intensity contrast. 
		(b) Light-field image is separated into foreground and background by using matrix factorization techniques.
		(c) A multi-view image array at a depth of 30 \si{um}. 
		(d-e) The central column and row extracted from a multi-view image array. Note that the view direction is changing vertically and horizontally. Such view changing accounts for the slope of lines in the EPI / phase-space. 
	}
	\label{Fig:LF_cells}
\end{figure}

The original LFM was designed by Levoy et al.~\cite{levoy2006light}, where a microlens array was inserted at the native image plane (NIP) of a widefield microscope to capture a four-dimensional (4D) light-field (including 2D lateral position and 2D angular information), as shown in Fig.~\ref{Fig:microlens-basedLF} and Fig.~\ref{Fig:LF_cells}. The angular information in turn relays depth information for volumetric reconstruction. In this way, LFM can capture volumetric information of the incident light in a single snapshot, allowing 3D imaging at video frame-rates. The ability of LFM to image neuronal activity over volumetric regions at video frame-rates is generating tremendous excitement in the neuroscience community with plenty of recent breakthroughs including the imaging of whole-brain neuronal activity in some small organisms (e.g. C. elegans, drosophila, larval zebrafish~\cite{prevedel2014simultaneous,aimon2019fast,cong2017rapid, nobauer2017video}). Despite its advantages in fast and large-scale 3D imaging, the original LFM has several critical limitations that in the beginning hampered its widespread use, such as compromised spatial resolution, presence of reconstruction artifacts, time-consuming reconstruction, low signal-to-noise ratio and image degradation caused by scattering in thick tissues.

In what follows, we review fundamental aspects of LFM, describe in Box~2 the wave-optics model used in this context~\cite{broxton2013wave}, and point out current challenges and viable solutions in Section~\ref{sec:Background}. Then, we briefly review the latest progress on LFM optical systems in Section~\ref{sec:OpticalSystems}, and present a thorough survey on state-of-the-art computational methods developed for improving LFM performance in Section~\ref{sec:ComputationalMethods}. We highlight the key aspects of these computational methods and also the potential gains that one can obtain when model-based priors are incorporated into data-driven approaches. We also discuss some pertinent suggestions on the application of varied machine learning concepts and techniques into LFM for neuroimaging in Section~\ref{sec:Discussion}. Finally, we conclude in Section~\ref{sec:Conclusion}.

\begin{framed}
	\label{Box:WaveModel}
	\singlespacing
	\mypar{Box 2: Light-field microscopy and its wave-optics model}

	\begin{center}
		\includegraphics[width = 13cm, 
		]{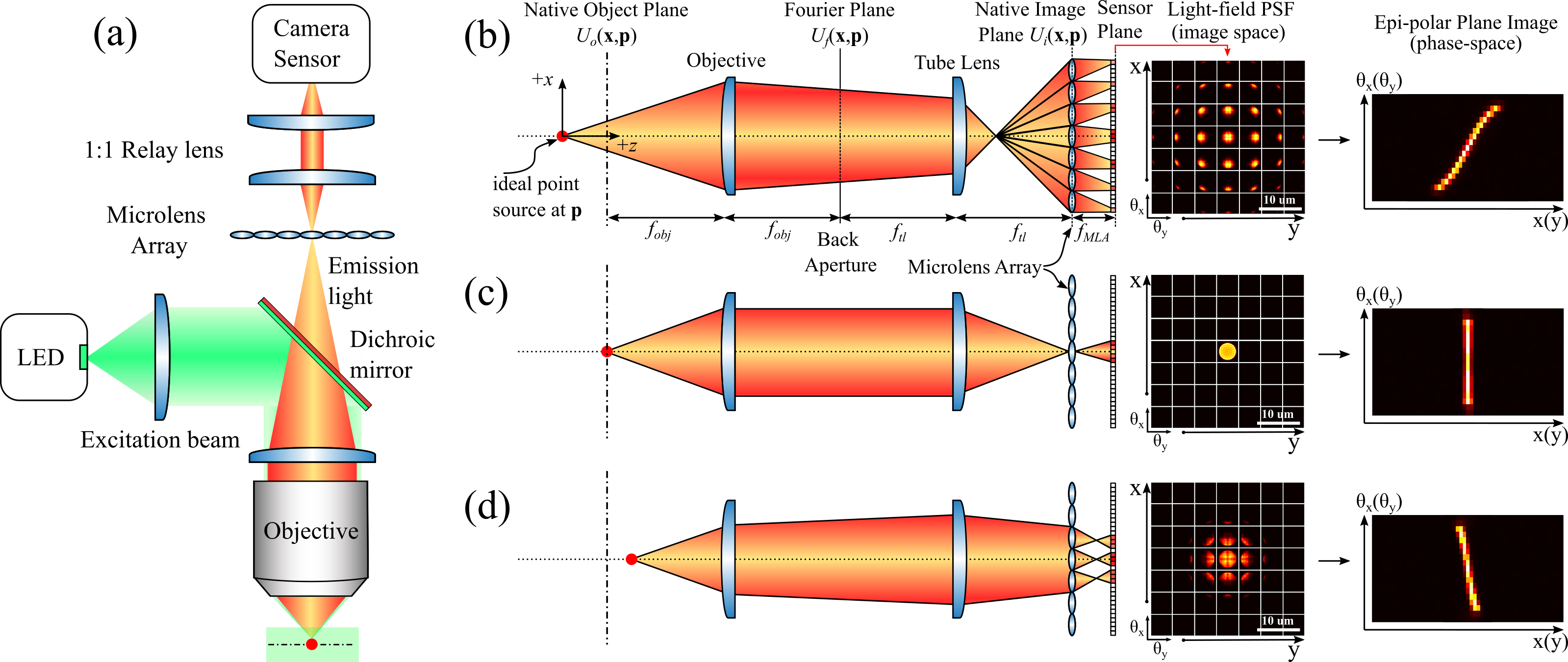} 
		\captionof{figure}{
			Optical model of the light-field microscope.
			(a) Original LFM configuration. A microlens array (MLA) is inserted at the native image plane (NIP) and the imaging sensor is placed at the back focal plane of the MLA~\cite{levoy2006light} to record 4D light-field. (b-d) LFM optical path and light-field patterns generated by one point source at different depths (i.e. axial positions). A point source below or above the native object plane, as shown in subfigure (b) and (d) generates a more complicated intensity pattern than a source in-focus as shown in subfigure (c), which indicates depth-dependent sampling. White lines in the LFM images depict the virtual profile of microlens in the image space and each square represents a sub-image associated with a specific lenslet. LFM simultaneously captures spatial and angular information which can be revealed in an Epi-polar Plane Image (EPI), a.k.a. phase-space or spatial-angular space. A deeper source leads to a more tilted line whose center indicates the lateral position.
		}
		\label{Fig:microlens-basedLF}
	\end{center}
	
	The microlens-based light-field imaging system in Fig.~\ref{Fig:microlens-basedLF} aims to transform the light-field from the world space into the image space of the main lens, thereby sampling the light-field at the sensor plane. Each lenslet with its underlying group of pixels forms an in-camera sampling scheme, analogous to a tiny camera with very few pixels, that observes the in-camera light-field. The diffraction pattern generated by an ideal point source when propagated through an optical system is the system's impulse response function, commonly referred to as the point spread function (PSF). A wave-optics forward model has been developed in~\cite{broxton2013wave} to model the light propagation from an ideal point source $\bp (p_1, p_2, p_3)$ to NIP~\eqref{Eq:LF_Ui}, passing through MLA~\eqref{Eq:LF_UMLA}, and finally arriving at the camera sensor~\eqref{Eq:LF_Uf}.
	\begin{equation} \label{Eq:LF_Ui}
	\small 
	\begin{split}
	&
	U_i(\bx,\bp)
	=
	\frac{M}{f_{obj}^2 \lambda^2} \exp\left(- \frac{i u}{4 \sin^2 (\alpha/2)} \right)
	\int_{0}^{\alpha} P(\theta) 
	J_o \left(v \frac{\sin (\theta)}{\sin (\alpha)} \right)
	\exp \left(-i u \frac{\sin^2 (\theta/2)}{2 \sin^2(\alpha/2)} \right)
	\sin(\theta) \dif \theta
	\end{split}
	\end{equation}
	\begin{equation}
	\small
	\label{Eq:LF_UMLA}
	U_{MLA}(\bx,\bp) =  U_i(\bx,\bp) \Phi(\bx) \,.
	\end{equation}
	\begin{equation}
	\small
	\label{Eq:LF_Uf}
	h(\bx,\bp) = \mathcal{F}^{-1} \{ \mathcal{F} \{ U_{MLA}(\bx,\bp) \} \cdot G(\hat{\bx}) \}
	\end{equation}
	\noindent
	Formulation~\eqref{Eq:LF_Ui} is derived from the scalar Debye theory.
	$J_o$ denotes the zero-th order Bessel function of the first kind. The auxiliary variables $v, u$ represent normalized radial and axial optical coordinates, defined as $v=k \sqrt{(x_1-p_1)^2 + (x_2-p_2)^2} \sin \alpha$ and $u=4 k n p_3 \sin^2 (\alpha/2)$, respectively, where $n$ denotes the refractive index of the material in which specimen is immersed, e.g. water or oil. Moreover, $\alpha = \sin^{-1} (\text{NA}/n)$ denotes the half-angle of the numerical aperture (NA); $k= 2 \pi /\lambda$ denotes the angular wavenumber. $P(\theta)$ denotes the apodization function of the microscope, e.g., $P(\theta) = \sqrt{\cos(\theta)}$ for Abbe-sine corrected objectives, $M = f_{tl} / f_{obj}$ denotes the magnification of the 4-f system, where $f_{tl}$ and $f_{obj}$ denote the focal lengths of the tube lens and the objective lens, respectively. Note, this equation only holds for low to moderate NA objectives.	
	In \eqref{Eq:LF_UMLA}, $\Phi(\bx)$ denotes the transmittance function, a.k.a. lens mask of the MLA which can be modeled by a convolution of the single lenslet transmittance $\phi(\bx)$ and 2D Dirac comb function $comb(\bx/d)$, leading to $\Phi(\bx) = \phi(\bx) \odot comb(\bx/d)$ where $\phi(\bx) = P(\bx) \exp \left(\frac{-ik}{2 f_{MLA}} \|\bx\|_2^2 \right)$ with pupil function $P(\bx)$, focal length $f_{MLA}$ and pitch $d$.
	In \eqref{Eq:LF_Uf}, $h(\bx, \bp)$ represents the PSF of the light-field microscope, $G(\hat{\bx})$ denotes the transfer function that propagates $U_{MLA}$ to the sensor plane, and $\hat{\bx}$ are spatial frequencies in the Fourier domain at the sensor plane. Under the paraxial assumption for this context, the Fresnel diffraction solution can be adopted to give $G(\hat{\bx}) = \exp(-\frac{i}{4\pi} \lambda f_{MLA} \|\hat{\bx}\|_2^2))$. Alternatively, for arbitrary distances between the MLA and the sensor, a more accurate Rayleigh-Sommerfeld diffraction solution can be used to predict $h(\bx,\bp)$~\cite{stefanoiu2019artifact}.
	\\
	\indent
	The light-field PSF has a complex and translation-variant pattern which depends on the specific 3D positions of the point source. Thus, the image formation cannot be modeled as a convolution of a scene with a corresponding PSF, as is commonly done in conventional image formation models. 
	Instead, the wavefront recorded at the sensor plane is described using a more general linear superposition integral~\eqref{Eq:point_source} and a corresponding discretized version~\eqref{Eq:LF_Forward_Model}.
	\begin{equation} \label{Eq:point_source}
	\small
	f(\bx) = \int |h(\bx,\bp)|^2 g(\bp) d \bp,
	\end{equation}
	\begin{equation} \label{Eq:LF_Forward_Model}
	\small
	\mathbf{f} = \mathbf{H} \mathbf{g}
	\end{equation}
	where $f(\bx)$ and $\mathbf{f}$ denote the continuous and discrete 2D intensity pattern at the sensor plane. $\bp \in \mathbb{R}^3$ is the position in a volume containing isotropic emitters whose combined intensities are distributed according to $g(\bp)$. 
\end{framed}

\vspace{-0.3cm}

\section{Background}
\label{sec:Background}

\vspace{-0.3cm}

Capturing the 4D light-field information with a single camera sensor introduces inherent trade-offs between spatial resolution and angular resolution because the camera pixels are now encoding 4 dimensions rather than 2. Under the application background of imaging neuronal activity, the reduced spatial resolution makes faithful reconstruction of the monitored volume from light-field data problematic, and this issue may become more severe due to light scattering in deep tissues which cannot be mitigated as in the case of raster-scanning, single-detector microscopes. This makes it more difficult to distinguish and localize neurons deep in the tissue. Moreover, some detailed neural structures, such as axons, dendrites, synapses, may not be clearly visualized from individual views. Accordingly, neural dynamics recorded from these fine neural structures may suffer from low signal-to-noise ratio~\cite{broxton2013wave,prevedel2014simultaneous,quicke2020subcellular}. Furthermore, the depth-dependent sampling leads to a non-uniform spatial resolution in the reconstructed volumes and can cause square-shaped artifacts at regions near the native object plane (NOP) due to coarse sampling. Moreover, the forward process projects a 3D object domain onto the 2D camera plane through convolution with a high-dimensional PSF that varies spatially in both lateral and axial dimensions. Traditional model-based volumetric reconstruction methods often exploit iterative deconvolution to solve this highly complex inverse problem, which aggravates the computational cost.

Facing these challenges, two parallel viable solutions can be exploited independently or jointly to address the above issues: (1) to improve the acquisition by designing alternative optical systems, or (2) to improve the reconstruction by designing advanced computational algorithms. 

With advanced optical systems design, LFM is endowed with desirable properties. For example, it is able to produce diverse sampling patterns by recoding the 4D information at the pupil plane instead of the native imaging plane, which improves information acquisition performance. This can be combined with refined selective illumination modalities, such as light-sheet excitation, to provide optical sectioning. These new features contribute to eliminating interference from background caused by out-of-focus light and scattering, thus increasing the signal-to-noise ratio, spatial resolution and volume coverage, which further fulfills the potential of LFM in imaging whole-brain neuronal activity on freely behaving organisms.

Aside from improving LFM optical systems, the other alternative is to improve post-processing performance by developing advanced computational methods capable of better exploiting angular information. Furthermore, proper learned or model-based priors can be incorporated in these methods. At present, we are witnessing an emerging trend that an increasing amount of effort is devoted to developing more efficient and effective computational algorithms for fast volumetric reconstruction, neuron localization and neuronal activity demixing using light-field data. These methods can be generally categorized into two classes: model-based approaches and data-driven approaches. The model-based category attempts to enhance the reconstruction by incorporating additional priors, such as smoothness, spatial and temporal sparsity, low-rank, depth-related shearing property and shift-invariance property in phase-space. In contrast, the data-driven a.k.a. learning-based category capitalizes on machine learning to pave the way to advanced computational LFM. In particular, more researchers are using deep learning approaches to achieve unprecedented performance. Also, the flexibility of the microscope system opens up the possibility for the creative combination of model-based methods with data-driven methods to improve the interpretability and credibility of learned models. 

\vspace{-0.2cm}

\section{Alternative optical systems for LFM}
\label{sec:OpticalSystems}

\vspace{-0.3cm}

Since the first design of LFM proposed in~\cite{levoy2006light}, there has been steady technological improvement on the optical system. A variety of advanced LFM systems have emerged to optimize the light-field recording~\cite{cong2017rapid,guo2019fourier,wang2019hybrid,zhang2020imaging}. These new designs are able to produce \textit{diverse sampling patterns} to improve acquisition performance~\cite{cong2017rapid,guo2019fourier}, or able to introduce \textit{refined illumination} strategies to provide optical sectioning~\cite{wang2019hybrid,zhang2020imaging}, or able to combine both benefits~\cite{wagner2019instantaneous}.

\begin{figure*}[tb]
	\begin{minipage}[b]{1\linewidth}
		\centering
		\includegraphics[width = 12cm, 
		]{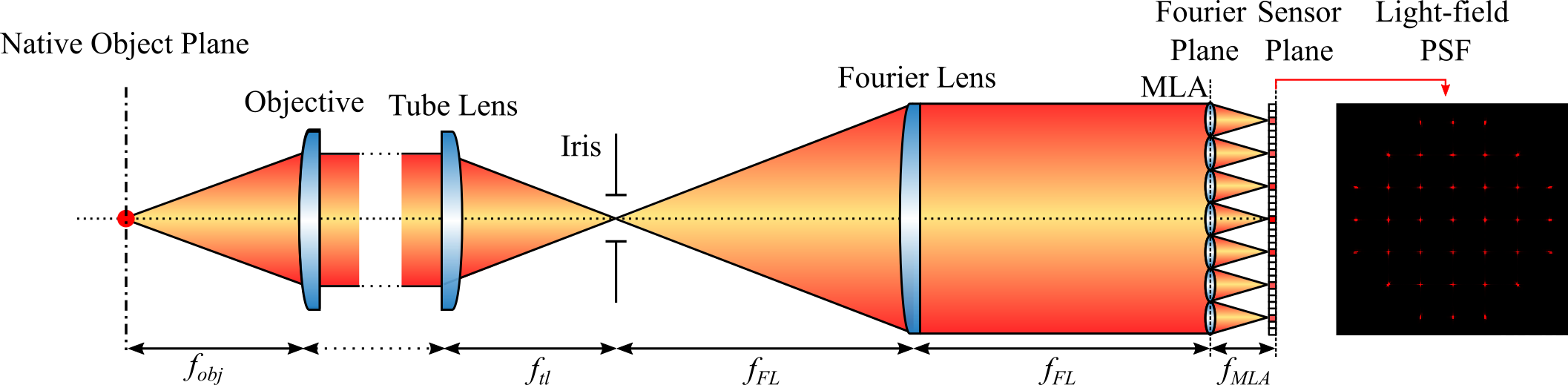}
		\\
		\small (a) Schematic of Fourier LFM~\cite{guo2019fourier}.
	\end{minipage}
	\\
	\vspace{+0.2cm}
	\\
	\begin{minipage}[b]{0.55\linewidth}
		\centering
		\includegraphics[width = 5cm, 
		trim=5cm 0cm 0cm 0cm,clip
		]{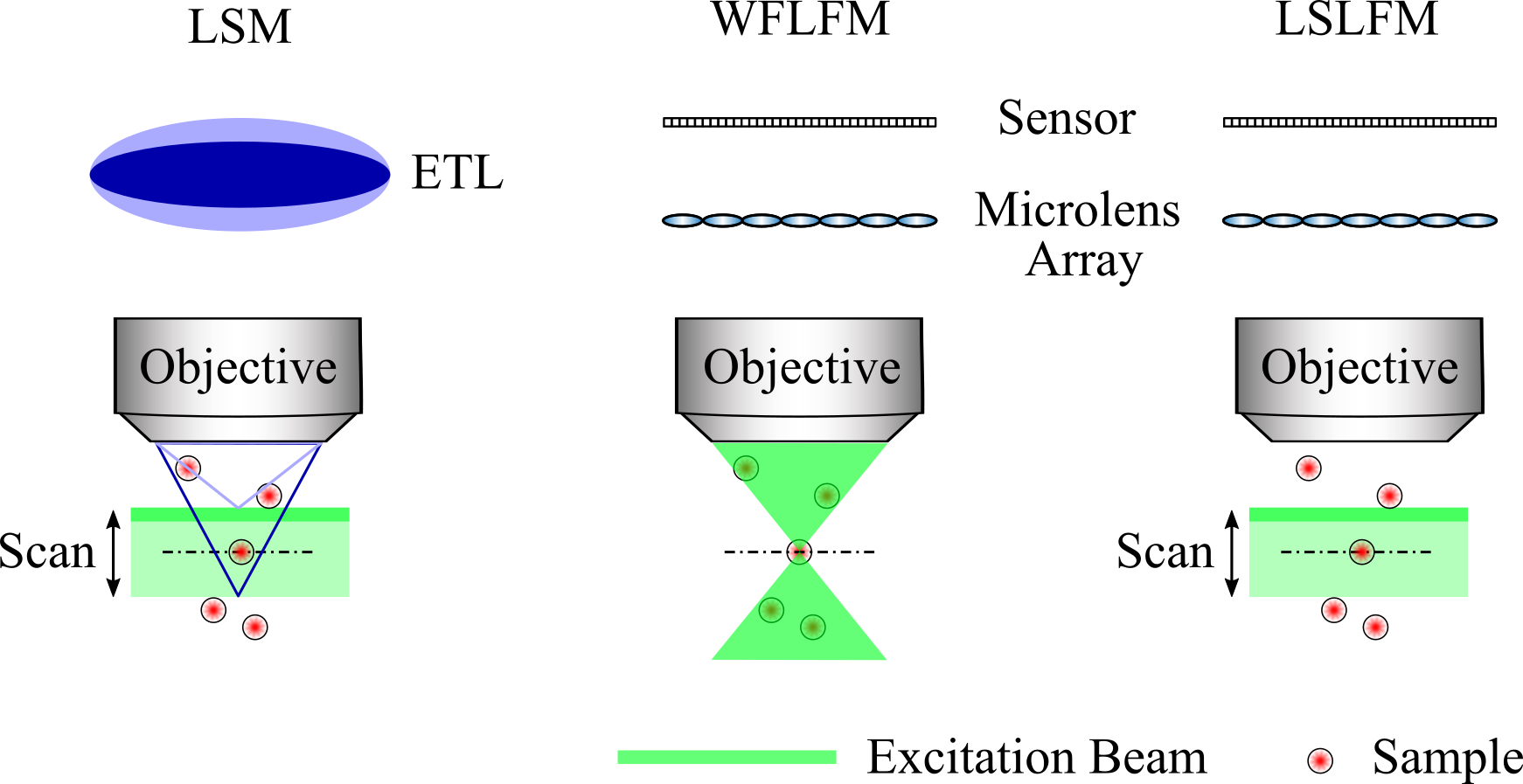}
	\end{minipage}
	\begin{minipage}[b]{0.44\linewidth}
		\centering
		\includegraphics[width = 5.5cm, 
		]{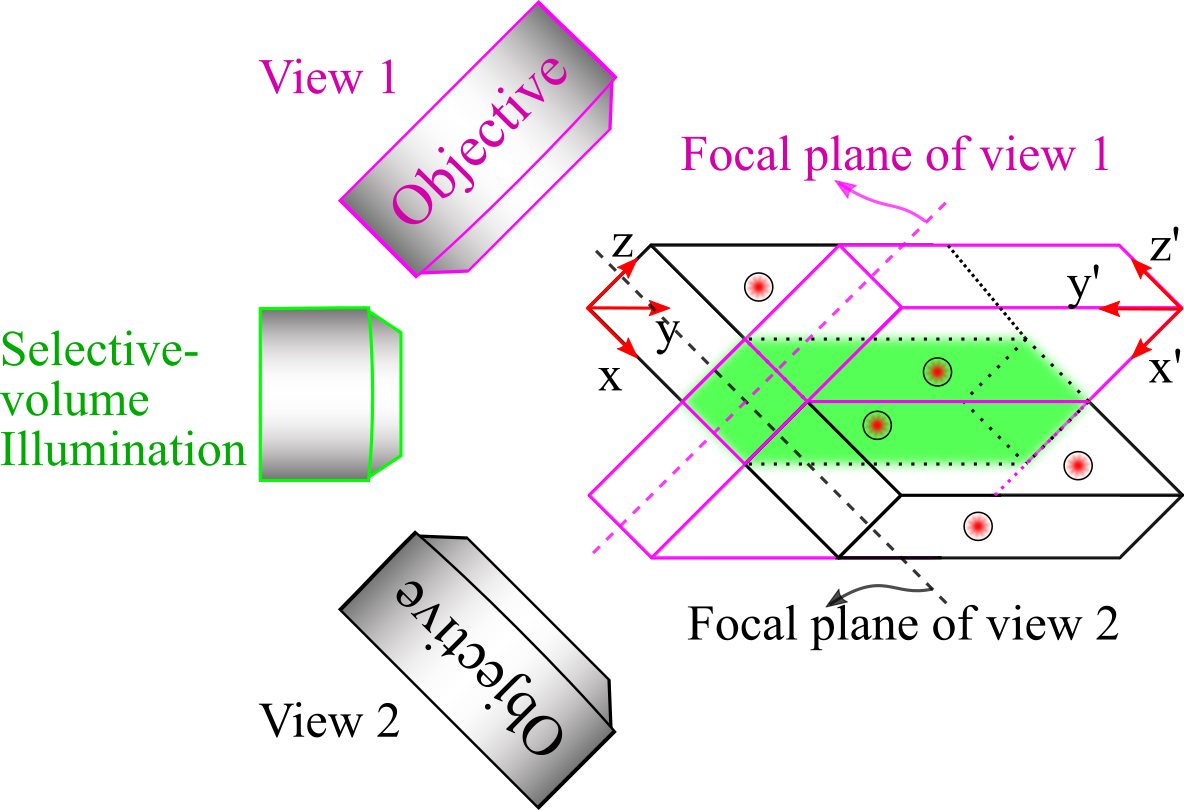}
	\end{minipage}
	\\
	\begin{minipage}[b]{0.55\linewidth}
		\centering
		\small (b) LFM with wide-field or light-sheet illumination~\cite{wang2019hybrid}.
	\end{minipage}
	\begin{minipage}[b]{0.44\linewidth}
		\centering
		\small (c) Schematic for Iso-LFM~\cite{wagner2019instantaneous}.
	\end{minipage}
	\captionof{figure}{Representative novel optical systems for LFM. (a) The Fourier LFM places a microlens array at the rear pupil plane of the objective, instead of the NIP. (b) LSLFM~\cite{wang2019hybrid} uses a scanning light-sheet for excitation and a microlens array for light-field imaging. (c) Isotropic spatial resolution light-field microscopy (Iso-LFM)~\cite{wagner2019instantaneous} is a dual-view light-field imaging system that combines selective-volume illumination with simultaneous acquisition of orthogonal (perpendicular) light-fields.}
	\label{Fig:NovelOpticalSystems}
\end{figure*}

Representative designs that exploit the insight of diverse sampling include eXtended field of view light-field microscopy (XLFM)~\cite{cong2017rapid} and Fourier light-field microscopy~\cite{guo2019fourier}, see also Fig.~\ref{Fig:NovelOpticalSystems}~(a). A common feature of these designs is that a (customized) microlens array is placed at the rear pupil plane of the imaging objective, instead of the NIP as in the original LFM which makes the position of the microlens array conjugated to the rear pupil plane. Such designs, in ideal conditions, can measure 2D spatially invariant point spread functions (PSF) which produce diverse sampling patterns and thereby avoid square-shaped artifacts near the focal plane.

Other works that leverage the insight of refined illumination include hybrid light-sheet and light-field microscopy (LSLFM)~\cite{wang2019hybrid} (see Fig.~\ref{Fig:NovelOpticalSystems}~(b)) and confocal light-field microscopy~\cite{zhang2020imaging}. Here, instead of using wide-field illumination, LSLFM~\cite{wang2019hybrid} uses a scanning light-sheet for excitation and a microlens array for light-field imaging. Such design simplifies the detection by limiting the illumination to the volume of interest. Based on the same optical design as XLFM, i.e. placing a microlens array at the objective's conjugate pupil plane for more diverse sampling, confocal LFM~\cite{zhang2020imaging} improves the illumination by shaping the excitation laser beam into a plane, which selectively illuminates an axial plane (x-z plane). This enables background-free, fast, volumetric imaging of neural dynamics deep in mouse and zebrafish brains.

Isotropic spatial resolution light-field microscopy (Iso-LFM)~\cite{wagner2019instantaneous} proposes a dual-view light-field imaging system to combine benefits of both
refined illumination and diverse sampling. In particular, Iso-LFM~\cite{wagner2019instantaneous} combines selective-volume illumination with simultaneous acquisition of orthogonal (perpendicular) light-fields to yield 3D images with high, isotropic spatial resolution and exhibits a significant reduction of reconstruction artifacts, thereby overcoming some current limitations of light-field microscopy implementations (see also Fig.~\ref{Fig:NovelOpticalSystems}~(c)). The selective-volume illumination is implemented by confining the excitation light spatially to the volume of interest, which optimizes signal-to-background contrast and minimizes erroneous reconstruction artifacts from out-of-volume emitters. Then, Iso-LFM detects the emitted fluorescence via two identical objectives placed perpendicular to each other and orthogonal to the illumination objective. This design provides a dual-view capability that enables dual-view data fusion and deconvolution of the simultaneously acquired light-fields. This configuration achieves effectively an isotropic spatial resolution of ~2 um, and also substantially reduces the presence of image planes containing reconstruction artifacts (so-called artifact planes).

\vspace{-0.3cm}

\section{Novel computational methods for LFM}
\label{sec:ComputationalMethods}

\vspace{-0.3cm}

Advanced optical hardware systems could introduce extra system complexity. Furthermore, changes to the optical system do not in themselves address the computational issues related to reconstructing volumes from light-fields. Therefore, to mitigate or overcome aforementioned issues, an increasing amount of research is devoted to advanced computational post-processing strategies. At present, we are witnessing an emerging trend of more efficient and effective computational algorithms being developed for fast volumetric reconstruction, neuron localization and signal demixing. 

In particular, some of these methods fall into the model-based category which advances the light-field model by incorporating additional priors, such as smoothness, non-negativity, spatial and temporal sparsity, low-rank, phase-space priors, etc.~\cite{prevedel2014simultaneous,nobauer2017video,stefanoiu2019artifact,verinaz2020volume,yoon2020sparse,lu2019phase,waller2012phase,liu20153d,pegard2016compressive,song20203D}. In contrast, others fall into the data-driven category which exploits machine learning to achieve more efficient computational solutions. In particular, by virtue of the superior modelling capability of deep neural networks in approximating highly-complex functions effectively, more and more algorithms based on deep learning are being proposed for fast and robust volumetric reconstruction and neuron localization~\cite{li2019deeplfm,wang2021real,wagner2021deep,verinaz2021deep,song2021Model-inspired}. These cutting-edge computational approaches have opened new avenues to push the limits of LFM. 

\vspace{-0.3cm}

\subsection{Model-based computational approaches}

\vspace{-0.3cm}

\subsubsection{Richardson-Lucy (R-L) deconvolution models and variations}

\vspace{-0.2cm}

Based on the pioneering light-field microscopy design~\cite{levoy2006light}, Broxton et al.~\cite{broxton2013wave} applied wave optics theory to model the light-field propagation (see Box~2) and proposed a Richardson-Lucy (R-L) 3D deconvolution algorithm for volumetric reconstruction from LFM images, which was further refined in~\cite{prevedel2014simultaneous}. 

However, as 3D deconvolution with light-field images is an ill-posed tomographic inverse problem~\cite{prevedel2014simultaneous}, wherein multiple different perspectives of a 3D volume are involved, the iterative R-L deconvolution method has high computational complexity and also tends to generate reconstruction artifacts, especially near the native focal plane. The underlying reason for these artifacts lies in the depth-dependent sampling of LFM; that is, LFM captures a different amount of information at different depths, which is reflected in the PSFs. In particular, the PSF corresponding to the in-focus plane (depth=0, at NOP) is highly redundant, implying very low angular resolution, and this leads to square-like artifacts during R-L deconvolution. To overcome these limitations, one feasible solution is to impose additional priors into the light-field model to constrain the reconstructed volume to lay in a more appropriate space. A series of studies described below have attempted to improve the efficiency and performance of volumetric reconstruction by incorporating advanced priors into the model.

Stefanoiu et al.~\cite{stefanoiu2019artifact} propose an aliasing-aware deconvolution method for artifact-free 3D reconstruction which employs depth-dependent anti-aliasing filters to remove artifacts from reconstructed volumes in each R-L iteration. Such iterative "deconvolution + anti-aliasing filtering" operations lead to the iterative aliasing-aware deconvolution, which enforces smooth priors in a smoothing expectation-maximization scheme.

Furthermore, Verinaz et al.~\cite{verinaz2020volume} incorporate more elaborated priors into the model to enforce more faithful solutions. To address the slow speed issue of the original R-L deconvolution, they also propose a method to simplify the light-field forward model which significantly reduces the computational complexity. Based on the simplified light-field model and elaborated priors, an Alternating Direction Method of Multipliers (ADMM) optimization strategy is developed to solve the 3D deconvolution problem and to achieve artifact-free volumetric reconstruction. 

\vspace{-0.3cm}

\subsubsection{Spatio-temporal factorization models}

\vspace{-0.3cm}

\begin{figure*}[tbp]
	\begin{center}
		\includegraphics[width = 18cm, 
		]{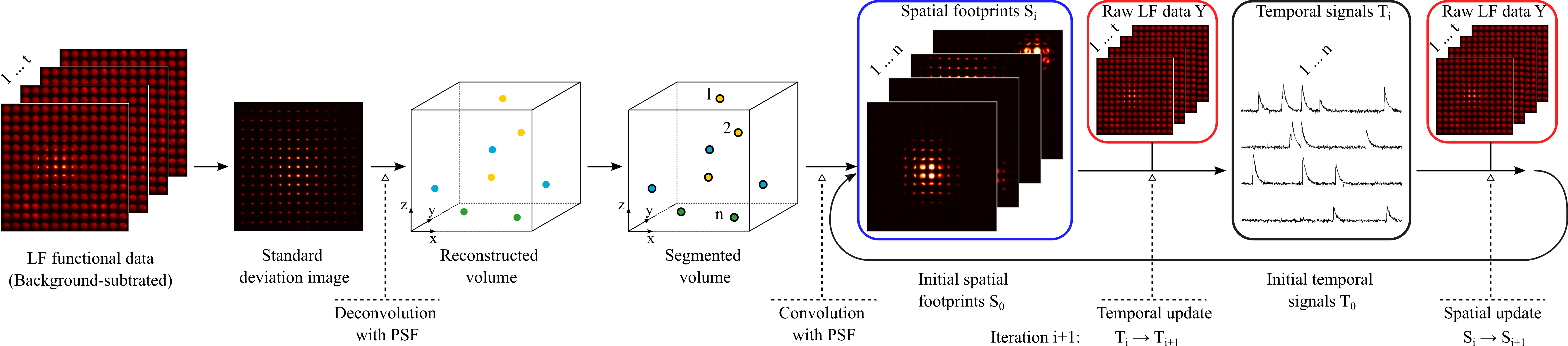}
		\captionof{figure}{Key concepts and processing flow of Seeded Iterative Demixing (SID) LFM. SID LFM performs localization by adding a segmentation operation to the deconvolved volume. The identified individual neurons result in footprints that aid the subsequent spatio-temporal demixing. Redrawn from~\cite{nobauer2017video}.
		}
		\label{Fig:SID_LFM.png}
	\end{center}
\end{figure*}

In certain scenarios, volumetric reconstruction is not the ultimate goal, as it may be of more interest to localize target neurons and extract functional activity. To identify target neurons and their activity, Nobauer et al.~\cite{nobauer2017video} proposed Seeded Iterative Demixing (SID) LFM which performs localization by adding a segmentation operation to the deconvolved volume. The identified individual neurons result in light-field footprints that aid the subsequent spatio-temporal demixing. Specifically, SID is an iterative source-extraction procedure for scattered LFM data that seeds inference with information obtained from remnant ballistic light. In this process, instead of frame-by-frame reconstruction of LFM images, SID achieves neuron localization and neuronal activity demixing by performing non-negative matrix factorization on scattered spatio-temporal (functional) LFM data. The key concepts of this method is illustrated pictorially in Fig.~\ref{Fig:SID_LFM.png}.

Considering that sparsity priors can improve reconstruction performance, spatial resolution and SNR, Yoon et al.~\cite{yoon2020sparse} propose a sparse decomposition LFM. This strategy converts the inherent temporal sparsity of neuronal activity into spatial sparsity of 2D images to achieve the level of resolution expected for sparse samples in densely packed samples. In particular, \cite{yoon2020sparse} decomposes the light-field time series into a low-rank non-negative component that corresponds to the static part, e.g. background, and a sparse non-negative component that corresponds to the neuronal activity. After the decomposition, high-resolution volume reconstruction can be achieved by applying Richardson-Lucy deconvolution with sparsity regularization to the sparse component rather than the raw LF images.

\subsubsection{Phase-space models}

Phase-space, a.k.a. spatial-angle space or epi-polar plane image (EPI) space~\cite{waller2012phase,bolles1987epipolar}, is an appropriate space to reveal the structure of light-field and this can be used to directly localize neurons and to obtain their functional activity. An example of EPI obtained from light-field microscopy images is depicted in Fig.~\ref{Fig:microlens-basedLF}. One particular useful property in phase-space is the depth-related shearing principle that each point in the volume traces out a tilted line in phase-space and the slope of this line is proportional to the depth. Such a property can also be explained using the phase-space Wigner function which states that the light propagation in space can be easily represented by a simple shearing operation in phase-space~\cite{waller2012phase}.

\mypar{Compressive LFM.}
Based on these principles, Liu et al.~\cite{liu20153d} demonstrates the use of phase-space imaging for 3D localization of multiple point sources inside scattering material. Pegard et al.~\cite{pegard2016compressive} further extend~\cite{liu20153d} by presenting a compressive light-field microscopy method which takes advantage of spatial and temporal sparsity of fluorescence signals to identify and localize each neuron in a 3D volume, with scattering and aberration effects naturally included and without ever reconstructing the volume. Specifically, non-negative matrix factorization is first used to sparsify the raw LFM video data. Once the training data is sparse, each frame can be separated into single sources using the phase-space sparse coding approach~\cite{liu20153d}. This leads to a "footprint" dictionary that is composed of each neuron's light-field signature. Then, each new LFM frame is decomposed as a linear, positive superposition of elements of the footprint dictionary. The coefficients of the sparse decomposition are a quantitative measure of functional fluorescence signals (in this case calcium transients), corresponding to the magnitude of the neuronal activity recorded.

\begin{figure*}[tbp]
	\begin{center}
		\includegraphics[width = 12cm, 
		]{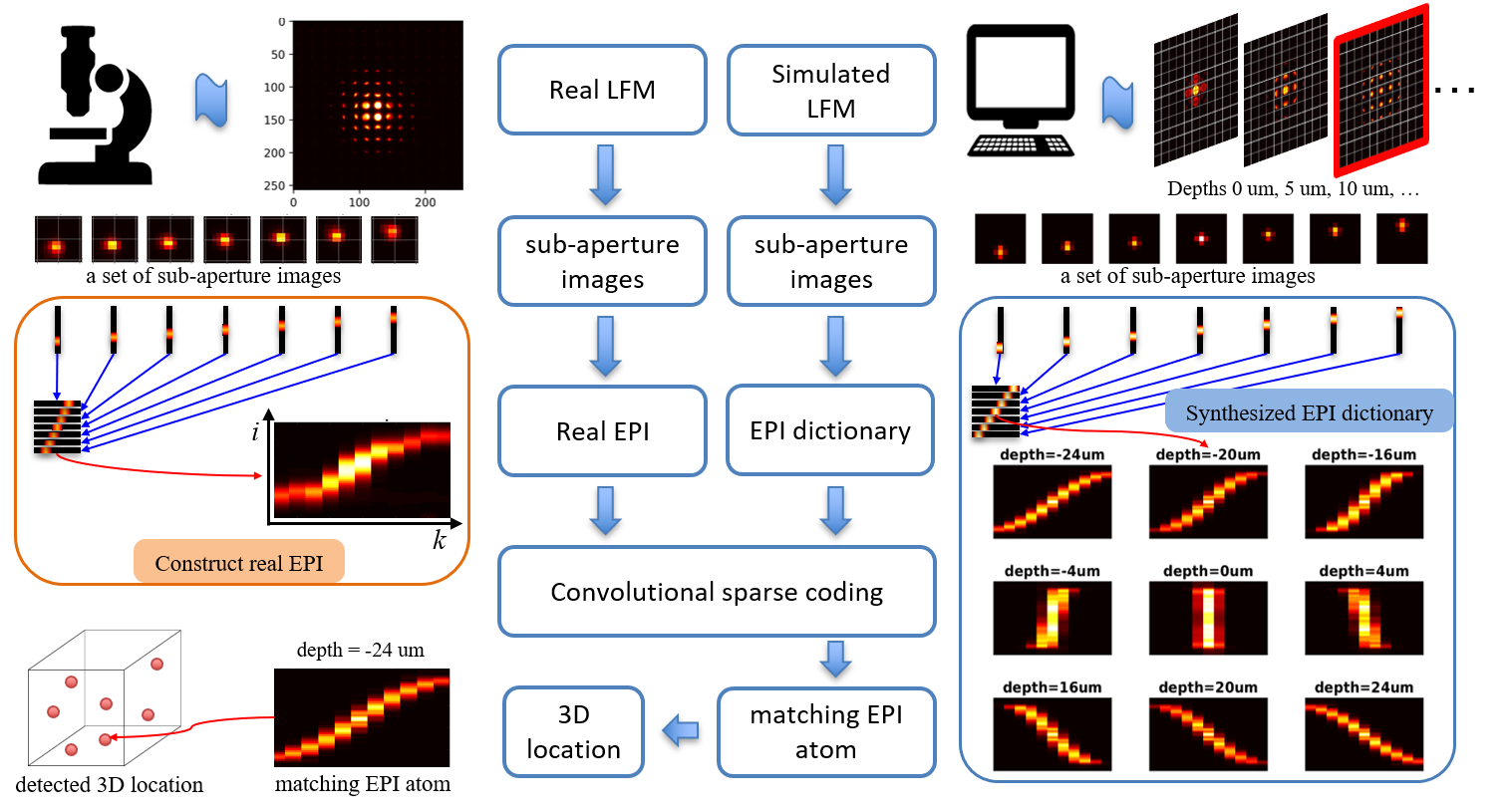}
		\captionof{figure}{Convolutional sparse coding LFM capitalizes on the shift-invariance property in the phase-space to perform convolutional sparse coding (CSC) on EPIs with respect to a depth-related EPI dictionary. The selected dictionary elements and corresponding sparse coefficients help to identify and localize neurons. Reproduced from~\cite{song20203D}.}
		\label{Fig:CSC_LFM}
	\end{center}
\end{figure*}

\mypar{Convolutional sparse coding LFM.}
Along this line, Song et al.~\cite{song20203D} propose a convolutional sparse coding LFM model, as shown in Fig~\ref{Fig:CSC_LFM}. Different from the compressive LFM~\cite{liu20153d,pegard2016compressive}, this work capitalizes on the shift-invariance property in the phase-space to perform convolutional sparse coding (CSC) on EPIs with respect to a depth-related EPI dictionary. The selected dictionary elements and corresponding sparse coefficients help to identify and localize neurons. The phase-space shift-invariance property brings two benefits: 1) It allows the convolution of a large input EPI with relatively small elements from an EPI dictionary, thus reducing the computational complexity; 2) It allows to only consider the depth range when synthesizing dictionary elements because the transverse shift can be revealed by a convolution operation. This reduces the dictionary size. More realistic and faithful dictionary design contributes to enhanced sparse coding performance, leading to improved robustness to light scattering and 3D localization accuracy. 

\vspace{-0.4cm}

\subsection{Data-driven computational approaches}

\vspace{-0.3cm}

In contrast to model-based methods, data-driven approaches attempt to automatically learn models from training data and incorporate discovered information into models, without requiring prior knowledge. This provides an appealing algorithmic alternative to overcome the shortcomings of model-based methods. Among data-driven approaches, deep neural networks, a.k.a. deep learning has attracted considerable attention as they provide unprecedented performance and efficiency in a variety of real-world signal and image processing tasks, including image de-noising, de-blurring, super-resolution, etc. To date, an explosive amount of effort is being invested into applying deep learning to the neuroimaging domain~\cite{li2019deeplfm,wang2021real,wagner2021deep,verinaz2021deep,song2021Model-inspired}. Multiple algorithms have recently been proposed to replace iterative 3D deconvolution by a deep neural network (e.g.~\cite{guo2020rapid}) to improve reconstruction speed and quality. We first introduce some representative purely data-driven approaches and then present some interpretable model-inspired data-driven approaches for LFM.

\vspace{-0.4cm}

\subsubsection{Purely data-driven approaches}

\vspace{-0.4cm}

\begin{figure*}[tbp]
	\begin{center}
		\includegraphics[width = 13cm, 
		]{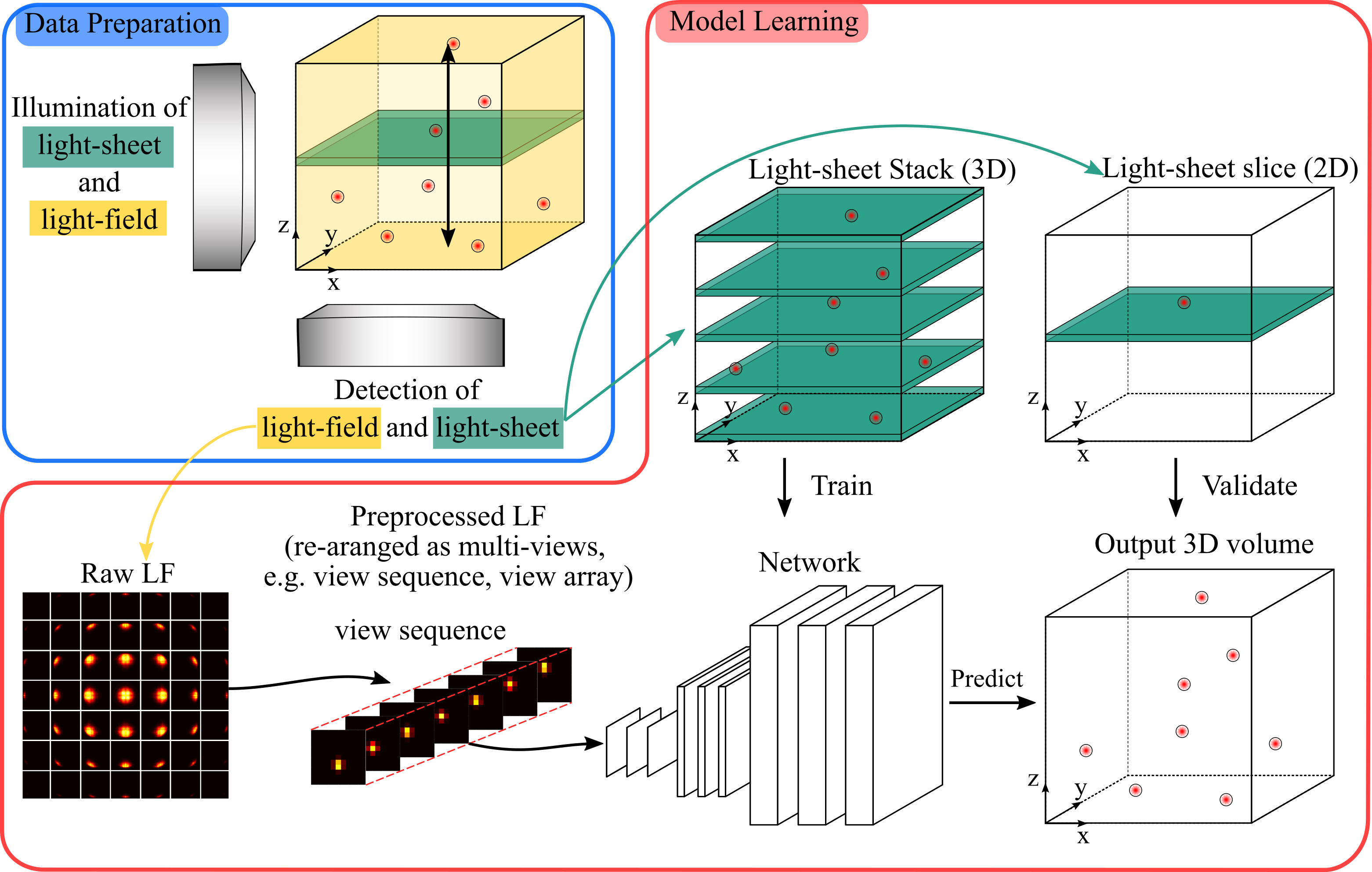}
		\captionof{figure}{Purely data-driven approaches, e.g. deep learning for light-field imaging, usually map a light-field image or its multi-view images to a volume after training on LFM images with another high-resolution image modality such as light-sheet microscopy images used as the ground truth. Redrawn from HyLFM-Net~\cite{wagner2021deep}.
		}
		\label{Fig:HyLFM-Net}
	\end{center}
\end{figure*}

\mypar{VCD-Net.}
Wang et al.~\cite{wang2021real} adapted U-net to achieve real-time and artifact-free volumetric reconstruction with uniform spatial resolution, referred to as view-channel-depth network (VCD-Net). Different from~\cite{li2019deeplfm} which performs R-L deconvolution on raw LFM images to obtain a low-resolution volume and then trains a 3D U-net for super-resolution, VCD-Net does not require the R-L deconvolution step. Instead, it extracts multiple views from raw light-field images and then directly maps them to the high-resolution volume (i.e. 3D image stack) through cascaded convolutional layers. To train the network, synthetic LFM images are generated by applying the forward model in~\cite{broxton2013wave} to ground truth high-resolution volumetric data acquired beforehand. In particular, the data preparation procedures include obtaining a number of high-resolution 3D volumes of stationary samples using synthetic or experimental methods and then performing light-field projection. Light-field projection converts these high-resolution 3D images into 2D light-field images which are then paired with ground truth data for network training. When the data is ready, VCD-Net is trained on the training dataset to transform input raw light-field back to 3D volumes, which would be compared with the high-resolution ground truth to guide optimization of the network. 

\mypar{LFM-Net.}
A similar concept was adopted in~\cite{page2021learning} to reconstruct high-quality confocal image stacks from LFM images using U-net which is trained on aligned pairs of LFM images and confocal image stacks to learn the direct mapping between them. Different from VCD-Net~\cite{wang2021real}, LFM-Net is trained on real LFM images rather than synthetic ones, and it does not convert raw 2D LFM images to a sequence of views in advance. During inference, LFM-Net achieves fast (10 frames per second, mostly delayed by the camera's exposure time of 100ms) and high quality (close to confocal imaging under similar optical settings) 3D volumetric reconstruction.

\mypar{HyLFM-Net.}
Although deep learning based-methods empirically demonstrate excellent image reconstruction performance and good generalization ability, no theoretical guarantees on generalization can be given. Out of this concern, Wagner et al.~\cite{wagner2021deep} suggest to extensively validate and, if needed, to retrain the network for each experimental setting. To this end, they developed a hybrid light-field light-sheet microscope, shown in Fig.~\ref{Fig:HyLFM-Net} where the light-field data is used as the input and the aligned high-resolution light-sheet data as ground truth to train the deep network. Moreover, when applying the trained network to new light-field data, new light-sheet data can be continuously acquired and this allows for fine-tuning or for retraining of the network on-the-fly if an inconsistency is found during continuous validation. Such a continuous validation mechanism serves as a dynamic calibration to ensure good generalization of the deep network. This is realized by adding a simultaneous, selective-plane illumination microscopy modality (e.g. light-sheet microscope) into the LFM setup which continuously produces high-resolution ground truth images of single planes for validation, training or refinement of the CNN. The training can thus be performed both on static sample volumes and dynamically from a single plane that sweeps through the volume during 3D image acquisition. 

\vspace{-0.3cm}
	
\subsubsection{Model-inspired data-driven approaches}

\vspace{-0.3cm}

Despite the good performance of modern deep networks in a variety of tasks, such purely data-driven approaches are also subject to some limitations~\cite{monga2021algorithm}. Apart from challenges that commonly appear in inverse bio-imaging problems, like imperfect knowledge of the forward model and lack of ground truth data, deep learning also has its own specific limitations. For example, generic deep networks are often empirically designed and typically adopt a hierarchical architecture composed of many layers and parameters. Although such designs endow deep networks tremendous capability of modelling obscure, even unknown physical systems, the network structures lack interpretability and involve excessive trainable parameters which may cause overfitting, degraded robustness and generalizability. Moreover, LFM for imaging neuronal activity has specific features, for example, the inherent spatial and temporal sparsity of fluorescence neuronal signals. This calls for development of novel deep learning methods that are able to fully exploit priors embedded in physical models and in the application considered. 

\begin{figure}[tbp]
	\begin{minipage}[b]{0.99\linewidth}
		\centering
		\includegraphics[width = 16cm, 
		]{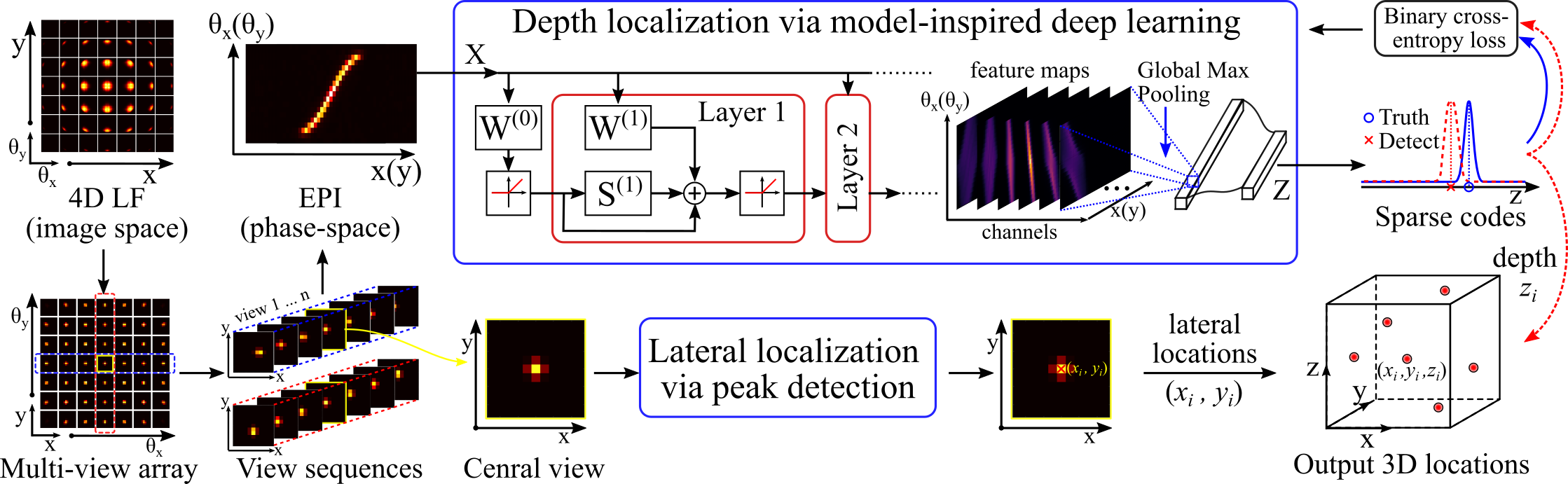}
		\\
		(a) CISTA-net LFM proposed in~\cite{song2021Model-inspired}.
	\end{minipage}
	\\
	
	\begin{minipage}[b]{0.99\linewidth}
		\centering
		\includegraphics[width = 13cm, 
		]{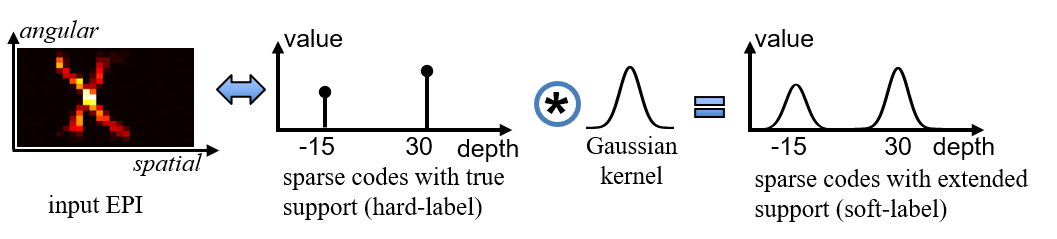}
		\\
		(b) Soft-label construction proposed in~\cite{song2021Model-inspired}.
	\end{minipage}
	\\
	\vspace{+0.1cm}

	\begin{minipage}[b]{1\linewidth}
		\centering 
		\includegraphics[width = 16cm
		]{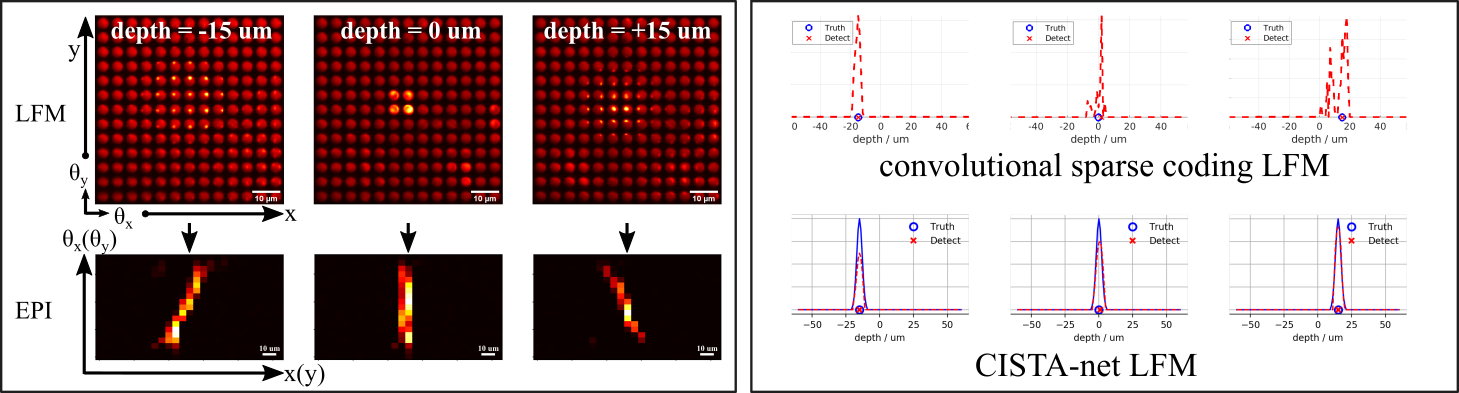} 
		\\
		(c) Predicted sparse codes and depth locations using model-based method~\cite{song20203D} and data-driven method~\cite{song2021Model-inspired}.
	\end{minipage} 	
	
	\caption{
			A model-inspired data-driven approach -- CISTA-net LFM~\cite{song2021Model-inspired} for neuron localization using LFM data. CISTA-net LFM combines deep learning with a phase-space convolutional sparse coding model to achieve fast and robust neuron localization. 
			(a) The architecture of the network is designed systematically by unrolling the convolutional Iterative Shrinkage and Thresholding Algorithm (CISTA) to efficiently solve a convolutional sparse coding (CSC) problem. The goal is to map an Epipolar Plane Image (EPI) to corresponding sparse codes that are associated with the 3D locations of target neurons. 
			(b) The knowledge distillation concepts are tailored for the task to train a compact model efficiently. In particular, after formulating the localization problem as a classification problem in the phase-space, the prior knowledge such as Gaussian-shape group structure and inter-class relationships are effectively incorporated into constructed soft-labels to achieve knowledge distillation from experts to the model. 		
			(c) Left: LFM images of a neuron at 3 different depths (-15, 0, +15 \si{um}) and corresponding EPIs without background (removed using matrix factorization). Right: Comparing sparse coding and depth detection performance of Convolutional sparse coding LFM~\cite{song20203D} and CISTA-net LFM~\cite{song2021Model-inspired}. It shows that data-driven approach~\cite{song2021Model-inspired} obtains sparse codes with higher quality and this leads to more accurate localization result.
	}
	\label{Fig:CISTA_net}
\end{figure}

\begin{figure*}[tbp]
	\begin{minipage}[b]{0.5\linewidth}
		\centering
		\includegraphics[width = 9cm, 
		]{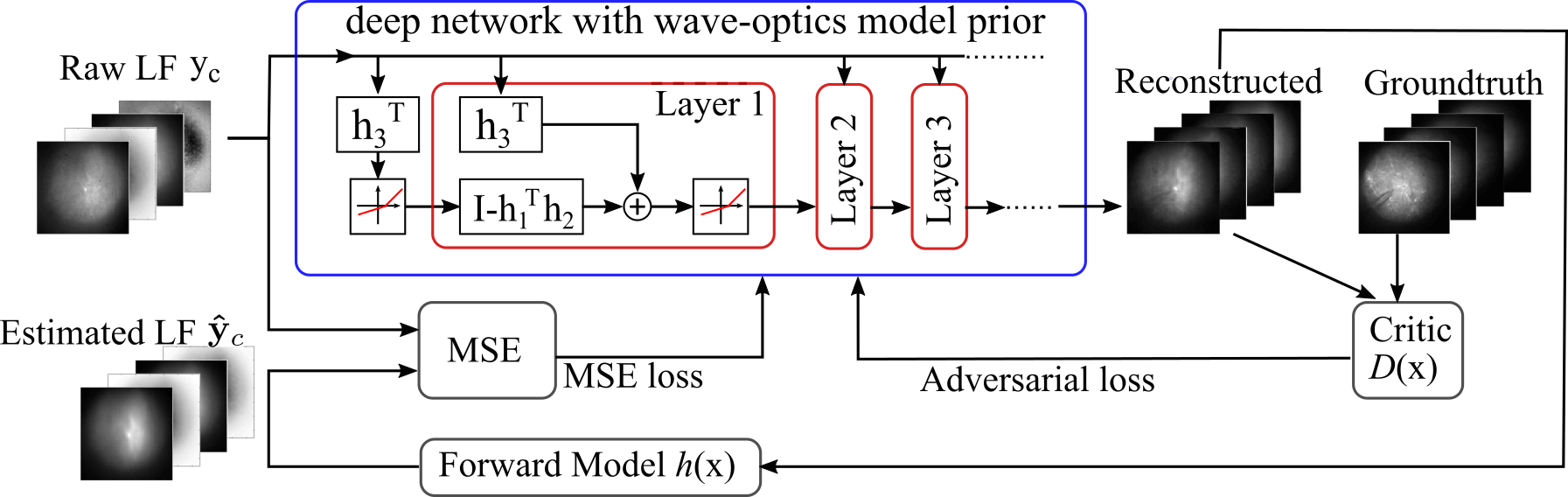}
		\\
		(a) LISTA-net LFM~\cite{verinaz2021deep}.
	\end{minipage}
	\begin{minipage}[b]{0.48\linewidth}
		\centering
		\includegraphics[width = 8cm, 
		]{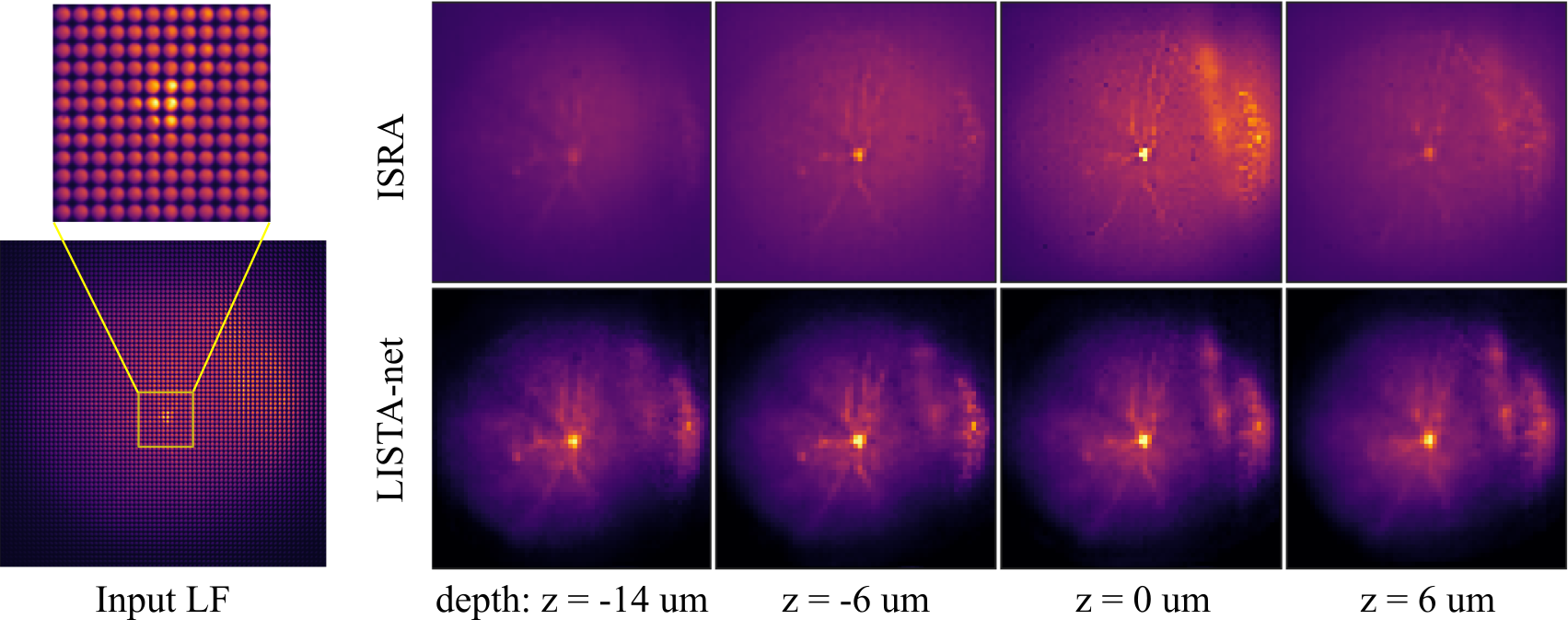}
		\\
		(b) Volumetric reconstruction results.
	\end{minipage}
	
	\caption{
		A model-inspired data-driven approach -- LISTA-net LFM~\cite{verinaz2021deep} for volumetric reconstruction using LFM data. LISTA-net LFM combines deep learning with light-field wave-optics model to achieve fast high-resolution 3D reconstruction using LFM data. 
		(a) The architecture of the network is designed by unrolling Iterative Shrinkage-Thresholding Algorithm (LISTA). An adversarial training strategy is exploited to train the network with unlabelled noisy measurements for unsupervised learning.
		(b) Reconstructed neurons from a LFM image using the improved R-L deconvolution method ISRA~\cite{prevedel2014simultaneous} (top) and data-driven approach --- LISTA-net LFM~\cite{verinaz2021deep} (bottom). It shows that the reconstruction from~\cite{verinaz2021deep} reveals details such as dendrites and axons more clearly and with less artifacts, as well as suppresses blurring and scattering at background. 
	}
	\label{Fig:LISTA_net}
\end{figure*}

\mypar{CISTA-net LFM.}
Song et al.~\cite{song2021Model-inspired} proposed a model-inspired deep learning approach to perform fast and robust 3D localization of neurons using light-field microscopy images. This is achieved by developing a deep network that efficiently solves a convolutional sparse coding (CSC) problem~\eqref{Eq:CSC_model} to map an EPI to corresponding sparse codes that are associated with the 3D locations of target neurons, as shown in Fig.~\ref{Fig:CISTA_net}:
\begin{equation}
\label{Eq:CSC_model}
\begin{array}{cl}
\underset{ \{\bz_m \}}{\min}
&
\frac{1}{2} \left\|\bX - \sum\limits_{m=1}^{M} \bd_m * \bz_m \right\|_2^2
+ \lambda \, \sum\limits_{m=1}^{M} \|\bz_m \|_1 \,.
\end{array}
\end{equation}
Here, $\bX$ denotes the EPI constructed from an observed LFM image, $\{ \bz_m \}_{m=1}^M$ denotes a series of sparse codes with respect to a EPI dictionary $\{ \bd_m \}_{m=1}^M$ with $M$ for the number of depths to be covered, $\| \cdot \|_2$ is the $\ell_2$ norm and $\| \cdot \|_1$ is the $\ell_1$ norm. The network architecture is designed systematically by unrolling the Convolutional Iterative Shrinkage and Thresholding Algorithm (CISTA) so that each iteration step of CISTA gives one layer of the network. The complete network is formed by concatenating multiple such layers. Therefore, forward passing through the network is equivalent to executing the CISTA algorithm a finite number of times. Note that the network contains domain specific layers which explicitly exploit physical priors, while the parameters are learned from a training dataset. Such design enables the network to leverage both domain knowledge implied in the model, as well as new knowledge learned from the data, thereby combining advantages of model-based and learning-based methods.

\mypar{LISTA-net LFM.}
Verinaz et al.~\cite{verinaz2021deep} proposed a novel deep network that combines the light-field wave-optics model with Learned Iterative Shrinkage-Thresholding Algorithm (LISTA), a well-known deep unrolling/unfolding network designed for efficient sparse coding, as shown in Fig.~\ref{Fig:LISTA_net} (a). Their analysis shows that when a LFM image is rearranged into a group of views presented as a 3D array, the LFM physics model can be conveniently approximated by a linear convolutional network. Accordingly, they modified LISTA to take a group of views as the input and to generate a high-resolution volume. At each unfolded iteration, the forward model and its transpose are computed repetitively to obtain a 3D volume at the output from the original light-field. It provides an effective way to incorporate physics knowledge into the network architecture, as well as improve its interpretability. Furthermore, inspired by Wasserstein Generative Adversarial Networks (WGANs), an adversarial training strategy is adopted, which makes it possible to train the network under realistic conditions such as lack of labelled data and noisy measurements, as only unlabelled data is needed to compute a properly designed adversarial loss. Fig.~\ref{Fig:LISTA_net} (b) shows reconstructed neurons compared with ISRA, an improved version of R-L deconvolution method. 

\vspace{-0.2cm}

\section{Discussion}
\label{sec:Discussion}

\vspace{-0.9cm}

\mypar{Ground truth for data-driven approaches.} 
When applying data-driven approaches to map light-field images to high-resolution 3D volumes, the ground truth can be given by other high-resolution imaging modalities including but not limited to confocal, light-sheet and multi-photon microscopy images~\cite{page2021learning,wagner2021deep}. However, due to high cost of obtaining ground truth, it is common that data-driven methods lack labelled data. This issue makes it challenging to directly exploit supervised learning methods for training. It may also cause overfitting and reduced generalization capabilities. To mitigate the impact of lack of ground truth data, other learning techniques and strategies, such as transfer learning, adversarial training, semi-supervised learning, unsupervised learning, can be adopted to develop feasible solutions. For example, one can use well-founded optic models to generate a large amount of synthetic data to train data-driven models~\cite{wang2021real,song2021Model-inspired}. The trained models can then serve as a good initialization for further fine-tuning on sparsely labelled real data. Alternatively, domain adaptation~\cite{tzeng2015simultaneous,tzeng2017adversarial,chadha2019improved} can be exploited to learn domain invariant feature representation via minimizing a distance metric, e.g. maximum mean discrepancy (MMD), or minimizing an adversarial loss between the source (synthetic) and target (real) distributions. In this way, a model trained on the source labelled data can then be directly applied to the target domain.

\mypar{Efficient data processing and model training.} 
Due to video-rate imaging speed, LFM can generate a large amount of data, which brings a severe challenge for real-time processing and analysing of neural dynamics. Fortunately, relying on much higher inference efficiency than iterative approaches, deep learning models are a promising avenue for real-time post-processing and analysis, such as volumetric reconstruction, 3D localization, neural activity demixing~\cite{wang2021real,page2021learning,wagner2021deep,verinaz2021deep,song2021Model-inspired}.

To train a compact model efficiently, knowledge distillation~\cite{hinton2015distilling,chen2017learning,yang2020distilling} concepts can be tailored for LFM based neuroimaging. Taking the depth localization of neurons using LFM for example, this problem can be formulated as a multi-class, multi-label classification problem after converting an original light-field into the Epipolar Plane Image (EPI), a particular spatio-angular feature in the phase-space~\cite{song2021Model-inspired}. Different from common classification tasks, this task has a specific feature that the hard-labels are well-structured. In particular, for an arbitrary class label, its adjacent/neighbouring class labels exhibit high correlation and coherence. This leads to a Gaussian-shape group structure that indicates the inter-class relationships, similar to the class probabilities represented by the softened teacher model's logits. Based on this observation, \cite{song2021Model-inspired} proposes to exploit the specific prior knowledge and directly construct the soft-labels (i.e. class probabilities) by performing a convolution between the hard-labels (i.e. ground truth) and a Gaussian kernel followed by normalization. In this way, the Gaussian kernel plays the role of a "temperature" scaling function as in knowledge distillation. It effectively smooths out the probability distribution to reveal inter-class relationships learned by human experts serving as a teacher model for the specific task. A larger kernel width corresponds to a higher "temperature" that allows the model to pay more attention to the inter-class correlations. Accordingly, the soft-labels are used to compute the loss function, more specifically, the distillation loss function. Soft-labels also bring some training benefits as incorporated inter-class relationships bring more guidance information to accelerate the training, as well as to enforce group sparsity into the network's prediction.

\vspace{-0.3cm}

\section{Conclusion}
\label{sec:Conclusion}

\vspace{-0.3cm}

In this article, we provided an extensive review to the latest progress on LFM for imaging neuronal activity. Advanced optical systems focus on improving the information acquisition performance via enhanced sampling diversity and refined illumination strategies, while advanced computational algorithms attempt to improve the post-processing performance via establishing more powerful models equipped with predefined or learned priors. Both strategies and their combinations promise to provide an unprecedented capability for imaging neuronal activity at a high resolution and speed across large volumes. We envision that more creative applications of deep learning and relevant data-driven approaches are being conceived and implemented for light-field microscopy. Along with computational algorithm becoming increasingly essential component of the post-processing, we will witness more uncovered potential and widespread usage of LFM.

\vspace{-0.2cm}

\section{Author Information}

\vspace{-0.6cm}

{	
	\singlespacing
	\small
	\noindent 
	\textbf{Pingfan Song} (ps898@cam.ac.uk) received his B.S. and M.S degree both from Harbin Institute of Technology (HIT), China, and his Ph.D. degree from University College London (UCL), UK. He was a research associate at Imperial College London, and is now a senior research associate at University of Cambridge. His research interests include signal/image processing, machine learning with applications on medical imaging, biological imaging, and other computational imaging tasks.
	
	\vspace{+0.1cm}
	
	\noindent
	\textbf{Herman Verinaz-Jadan} (herman.verinaz-jadan17@imperial.ac.uk) received his B.S. degree from Escuela Superior Politecnica del Litoral (ESPOL), Ecuador, and his M.S. degree from Imperial College London, UK. He is currently working toward the PhD degree in Imperial College London, UK. His research interests include sparsity-driven signal/image processing, machine learning with applications in the solution of inverse problems, Light Field Microscopy. 
	
	\vspace{+0.1cm}
	
	\noindent
	\textbf{Carmel L. Howe} (carmel.howe@imperial.ac.uk) received her MEng and Ph.D. degrees in Electrical and Electronic Engineering from the University of Nottingham in 2014 and 2018, respectively. She is a research associate at Imperial College London, UK. Her research combines the fields of neurophysiology, optical engineering, signal and image processing.
	
	\vspace{+0.1cm}
	
	\noindent
	\textbf{Amanda J. Foust} (a.foust@imperial.ac.uk) studied Neuroscience with emphasis in computation and electrical engineering (BSc) at Washington State University, and Neuroscience (MPhil, PhD) at Yale University. She is a Royal Academy of Engineering Research Fellow and Lecturer in the Imperial College London. The aim of her research programme is to engineer bridges between cutting-edge optical technologies and neuroscientists to acquire new, ground-breaking data on how brain circuits wire, process, and store information.
	
	\vspace{+0.1cm}
	
	\noindent
	\textbf{Pier Luigi Dragotti} (Fellow 2017) (p.dragotti@imperial.ac.uk) is Professor of Signal Processing in the Electrical and Electronic Engineering Department at Imperial College London. He has held several visiting positions at Stanford University, in 1996, at Bell Labs, Lucent Technologies, in 2000 and at Massachusetts Institute of Technology (MIT) in 2011. He was Technical Co-Chair for the European Signal Processing Conference in 2012, Associate Editor of the IEEE Transactions on Image Processing from 2006 to 2009 and Editor-in-Chief of the IEEE Transactions on Signal Processing (2018 to 2020). He was also the recipient of an ERC starting investigator award for the project RecoSamp. Currently, he is IEEE SPS Distinguished Lecturer. His research interests include sampling theory, wavelet theory and its applications, sparsity-driven signal processing with application in image super-resolution and neuroscience.
	
	\vspace{+0.1cm}
	
	\phantom{??}
}

{
	\singlespacing
	\small 
	\bibliographystyle{IEEEtran}
	\bibliography{mybib_LFM}
}


%

\end{document}